\documentclass[aps,twocolumn,showpacs,pra,amsmath,amssymb,floatfix,superscriptaddress]{revtex4-1}

\usepackage{color}
\usepackage{bbm}
\usepackage{graphicx}
\usepackage{dcolumn}
\usepackage{bm}
\usepackage{array}
\usepackage{float}
\usepackage{supertabular}
\usepackage{longtable}
\usepackage{mathrsfs}
\usepackage{txfonts}
\usepackage{wasysym}
\usepackage[T1]{fontenc}
\usepackage[latin1]{inputenc}
\usepackage[usenames,dvipsnames]{xcolor}
\usepackage{amsmath}
\usepackage[colorlinks=true,citecolor=Cerulean,linkcolor=RubineRed,urlcolor=Cerulean]{hyperref}

\begin{document}

\title{Optimal control of superconducting gmon qubits using Pontryagin's minimum principle: preparing a maximally entangled state with singular bang-bang protocols }

\author{Seraph Bao}
\affiliation{Department of Physics and Astronomy, University of British Columbia, Vancouver, British Columbia, Canada V6T 1Z4}
\author{Silken Kleer}
\affiliation{Department of Physics and Astronomy, University of British Columbia, Vancouver, British Columbia, Canada V6T 1Z4}
\author{Ruoyu Wang}
\affiliation{Department of Physics and Astronomy, University of British Columbia, Vancouver, British Columbia, Canada V6T 1Z4}
\author{Armin Rahmani}
\affiliation{Department of Physics and Astronomy, Western Washington University, Bellingham, Washington 98225, USA}

\date{\today}

\begin{abstract}

We apply the theory of optimal control to the dynamics of two ``gmon'' qubits, with the goal of preparing a desired entangled ground state from an initial unentangled one. Given an initial state, a target state, and a Hamiltonian with a set of permissible controls, can we reach the target state with coherent quantum evolution and, in that case, what is the minimum time required? The adiabatic theorem provides a far from optimal solution in the presence of a spectral gap. Optimal control yields the fastest possible way of reaching the target state and helps identify unreachable states. In the context of a simple quantum system, we provide examples of both reachable and unreachable target ground states and show that the unreachability is due to a symmetry. We find the optimal protocol in the reachable case using three different approaches: (i) a brute-force numerical minimization (ii) an efficient numerical minimization using the bang-bang ansatz expected from the Pontryagin minimum principle, and (iii) direct solution of the Pontryagin boundary value problem, which yields an analytical understanding of the numerically obtained optimal protocols. Interestingly, our system provides an example of singular control, where the Pontryagin theorem does not guarantee bang-bang protocols. Nevertheless, all three approaches give the same bang-bang protocol.

\end{abstract}
 
\maketitle

\section{Introduction}

The coherence times of quantum devices are rapidly increasing~\cite{Dowling03,Paladino14}, promising novel quantum machines and technologies [see, e.g., Refs. (\cite{Monroe13, Devoret13, Awschalom13, Mohseni17})]. Optimal control plays a crucial role in driving machines governed by classical laws of physics, enhancing their performance and efficiency. Given the finite coherence times of quantum devices, optimal control may be even more important for quantum technologies, as slow performance may make certain coherent processes altogether impossible. How can we optimally manipulate systems and devices governed by coherent quantum dynamics (see Ref.~\cite{Glasser15} for a review)? What are the characteristics of optimal quantum control protocols? 

Despite a long history, especially in the few-body context~\cite{Peirce88,Kosloff89} (see also Refs.~\cite{Werschnik07,Brif10}), several practical and fundamental questions remain unanswered~\cite{Rahmani13a,Glasser15}. With numerous novel applications to many-body dynamics~\cite{Rahmani11,Doria11,Caneva11}, cold atoms~\cite{Hohenester07,Rohringer08,Goerz11}, and quantum information processing (e.g., topological quantum computing~\cite{Karzig15a, Karzig15, Rahmani17} and variational quantum algorithms~\cite{Peruzzo14,Wecker15,Wecker16,	McClean16,Omalley16,Farhi,Yang}), quantum optimal control has emerged as an exciting frontier in nonequilibrium quantum dynamics. The objectives of quantum optimal control are diverse. We may want to steer the quantum states to a certain desired target state from a fixed initial state, prepare states with certain figures of merit (e.g., squeezed states)~\cite{Grond09,Pichler15}, cool down the quantum systems~\cite{Sklarz04,Hoffmann11,Choi11,Choi12,Rahmani13}, or generate a unitary evolution operator (e.g., a quantum gate) independently of the initial state~\cite{Tesch01,Palao02, Palao03,Sporl07,Montangero07,Schutjens13,Zahedinejad15}.

One particular application of optimal control is finding shortcuts~\cite{Torrontegui} to the adiabatic evolution without any modification to the form of the Hamiltonian: starting from the ground of a Hamiltonian (for certain values of the coupling constants), how should we change these tunable coupling constants, within a permissible range, to reach another ground state (corresponding to different values of the coupling constants) as fast as possible? Constraining the range of the coupling constants is one (not unique but experimentally motivated) way of fixing the energy scale of the Hamiltonian, which is important for making the problem well defined (an unphysical arbitrary increase of the energy scale can make all processes arbitrarily fast). Even in this simplest case, many questions remain unanswered. Of particular interest is the shortest possible time to reach the target state from a given initial state. This time scale sets a permissible-Hamiltonian-dependent measure of \textit{distance} between the initial and the target state, the properties of which are relatively unexplored.

Focusing on a simple highly tunable two-qubit system~\cite{Chen14} relevant to new superconducting devices, here we explore the properties of optimal control for transforming the quantum state from a given initial state. We consider, as an example, the creation of an entangled singlet state. Due to the purely quantum nature of these states and their sensitivity to environmental perturbations, it is difficult to prepare them directly. In order to prepare entangled states, one typically initiates the quantum system in an easy-to-create direct product state and uses quantum evolution to transform the state to an entangled one. A simple method for such state transformation is based on the quantum adiabatic theorem: Using a tunable device the Hamiltonian of which (for different parameters) supports both trivial and entangled ground states, we can reach the entangled state by slowly changing the device parameters. Here, we are interested in creating the entangled state in the context of shortcuts to adiabaticity. Optimal control has also been applied to the creation of two-qubit entangled states in the general context of perfect entanglers, i.e., unitary operators that map various direct-product states to the entangled Bell states~\cite{Goerz15a,Goerz15b,Goerz17}.

According to the adiabatic theorem, preparation of the target ground state is achieved once the process takes much longer than a characteristic time scale set by the energy gap between the ground state and the first excited state. The long time scales required by the adiabatic theorem are undesirable. As only the final state is of interest, we do not need to constrain the system to remain in the instantaneous ground state, making these time scales unnecessary. This is the essence of the optimal-control approach to finding shortcuts to adiabaticity. In the absence of this constraint, can we reach the target ground state exactly in a finite time? In that case, what is the best way of changing the Hamiltonian parameters, i.e., the optimal protocol. What is the shortest time required?

Here we address these questions using a two-step approach. We choose a measure of distance (based on the wavefunction overlap) between the final and the target states. For a given total time, we find the protocol which minimizes this distance. We then keep increasing the total time until the optimal distance vanishes. Our permissible Hamiltonians are characterized by two bounded control knobs. \textit{A priori}, our scheme is not concerned with a trajectory on the ground-state manifold and the system can be arbitrarily excited with respect to the instantaneous ground state. Interestingly, we find that in a case where the adiabatic transformation fails due to a level crossing, controlled nonadiabatic dynamics is also incapable of preparing the target state. While optimal control relies on nonadiabatic dynamics and should be naively insensitive to the properties of an adiabatic trajectory between the initial and target states, the same symmetry that protects a level crossing and prevents an adiabatic passage, forbids the more general transformation by nonadiabatic evolution.

In another case where the two ground states are not separated by a level crossing, we find that our optimal nonadiabatic protocol prepares the target state exactly with a sequence of square pulses, known as a bang-bang protocol. The general problem of finding such protocols is of considerable interest (particularly in the many-body context) and as we see in this simple model the knowledge of the bang-bang form of the protocol may significantly reduce the computational complexity of the problem. While, generically, bang-bang protocols are expected from the Pontryagin theorem~\cite{pontry1,pontry2}, we can have singular controls that may not be bang-bang. Interestingly, in our model, we do find a singular interval. Nevertheless, the optimal solution turns out to be bang-bang.

\section{Model and setup}

Consider a system described by a Hamiltonian with tunable parameters and an initial state that is the ground state of this Hamiltonian for certain values of these parameters. When attempting to transform this state by Hamiltonian evolution, the desired final state may be reachable or unreachable. In the special case where the initial and final states are both ground states of a gapped Hamiltonian, the adiabatic theorem implies that the desired state can be reached at least in the infinite time limit.

Consider as an example the case of  preparing the maximally entangled singlet state of two qubits:
\begin{equation}\label{eq:target}
  |\psi_{\rm target}\rangle={1\over \sqrt{2}}(|\uparrow\downarrow\rangle-|\downarrow\uparrow\rangle),
\end{equation}
from an unentangled initial ground state. The up and down spins are eigenstates of $\sigma^z$ [$\sigma^z|\uparrow\rangle=|\uparrow\rangle$ and $\sigma^z|\downarrow\rangle=-|\downarrow\rangle$], where $\sigma^{x,y,z}$ denote the Pauli matrices. 
The form of the Hamiltonian is set by the architecture of the device. Motivated by the coupling between powerful gmon qubits developed in the Martinis group (see Chen \textit{et al.}~\cite{Chen14}), we choose
\begin{equation}\label{eq:hamil}
H=B_1\sigma^x_1+B_2\sigma_2^x+
J\left(\sigma^x_1\sigma_2^x+\sigma^y_1\sigma_2^y\right).
\end{equation}

The gmon qubits allow for much more control. For two qubits, we can add other single-qubit terms $\sigma^y_{1,2}$ and $\sigma^z_{1,2}$. It is also possible to generate an effective $\sigma^z_1\sigma_2^z$ interaction through virtual tunneling to higher levels outside the qubit sector (it is possible to limit leaking outside the qubit manifold~\cite{Martinis}). The larger the number of control knobs, the more power we have in state transformation. However, finding optimal controls becomes more complicated with more control fields. In this paper we focus on the simplest case, where only changing two parameters in time generates the dynamics. This simple case is illuminating from a theoretical perspective. It also provides a fast and robust way of creating an entangled state.

The above Hamiltonian has three parameters. To restrict ourselves to only two tuning parameters, we focus on two cases with $\pm B_1= B_2=B$  for simplicity. We assume that both $B$ and $J$ parameters can be tuned as a function of time in the following range:
\begin{equation}\label{eq:range}
0\leqslant  B(t),J(t) \leqslant \Lambda,
\end{equation}
where we set $\Lambda=1$ (we have also set $\hbar$ to unity). Experimentally, the coupling can be tuned within a range range ${J\over 2\pi}\sim 10^3$ MHz~\cite{Chen14}. We also note that the parameters in the Hamiltonian are tuned indirectly. The coupling $J$, e.g., depends on the inductances of linear inductors connecting the qubits to the ground, the inductance of a Josephson junction between qubits, and the resonance frequency of the qubits \cite{Chen14}. The Josephson inductance, in turn, depends on a phase difference that can be tuned by applying a dc flux. The dependence of $J$ on the flux is calibrated through a simple mapping. The discussion becomes more transparent if we work with the effective Hamiltonian parameters $B$ and $J$ instead of the physical parameters such as the flux..

Note that the target state~\eqref{eq:target} is the ground state of the Hamiltonian~\eqref{eq:hamil} for $J=1$ and $B=0$. For $J=0$, the two qubits are decoupled and the ground states for the two cases $\pm B_1= B_2=B$ are unentangled direct products
\begin{eqnarray}
  |\psi^+(0)\rangle &=& {1\over 2}(|\uparrow\uparrow\rangle-|\uparrow\downarrow\rangle-|\downarrow\uparrow\rangle+|\downarrow\downarrow\rangle), \\
  |\psi^-(0)\rangle &=& {1\over 2}(|\uparrow\uparrow\rangle-|\uparrow\downarrow\rangle+|\downarrow\uparrow\rangle-|\downarrow\downarrow\rangle).\label{eq:psi-}
\end{eqnarray}

We comment that the above set of tunable parameters and the initial states are chosen for a nontrivial connection to shortcuts to adiabaticity. The speed limit we find is specific to the permissible Hamiltonian form and the chosen initial state, and can be viewed as a system-dependent minimal time of transforming our initial states to the target state. If the experimental goal is to merely create the singlet target state \eqref{eq:target}, other unentangled initial states and control parameters may be more convenient. For instance, an initial state $|\uparrow\downarrow\rangle$ can be rotated to an entangled superposition ${1\over\sqrt{2}}(|\uparrow\downarrow\rangle+i|\downarrow\uparrow\rangle)$, by only using a rotation, $e^{i{\pi\over 4}\tau_x}$,  generated by the coupling operator $\tau_x\equiv\left(\sigma^x_1\sigma_2^x+\sigma^y_1\sigma_2^y\right)/2$, with $\tau_x|\uparrow\downarrow\rangle=|\downarrow\uparrow\rangle$ and $\tau_x|\downarrow\uparrow\rangle=|\uparrow\downarrow\rangle$.
 This entangled state can then be transformed to the target state \eqref{eq:target} by applying a field $B_z\sigma_2^z$ for a time $\pi\over 4B_2^z$.

\begin{figure}[]
\centerline{\includegraphics[width=\columnwidth]{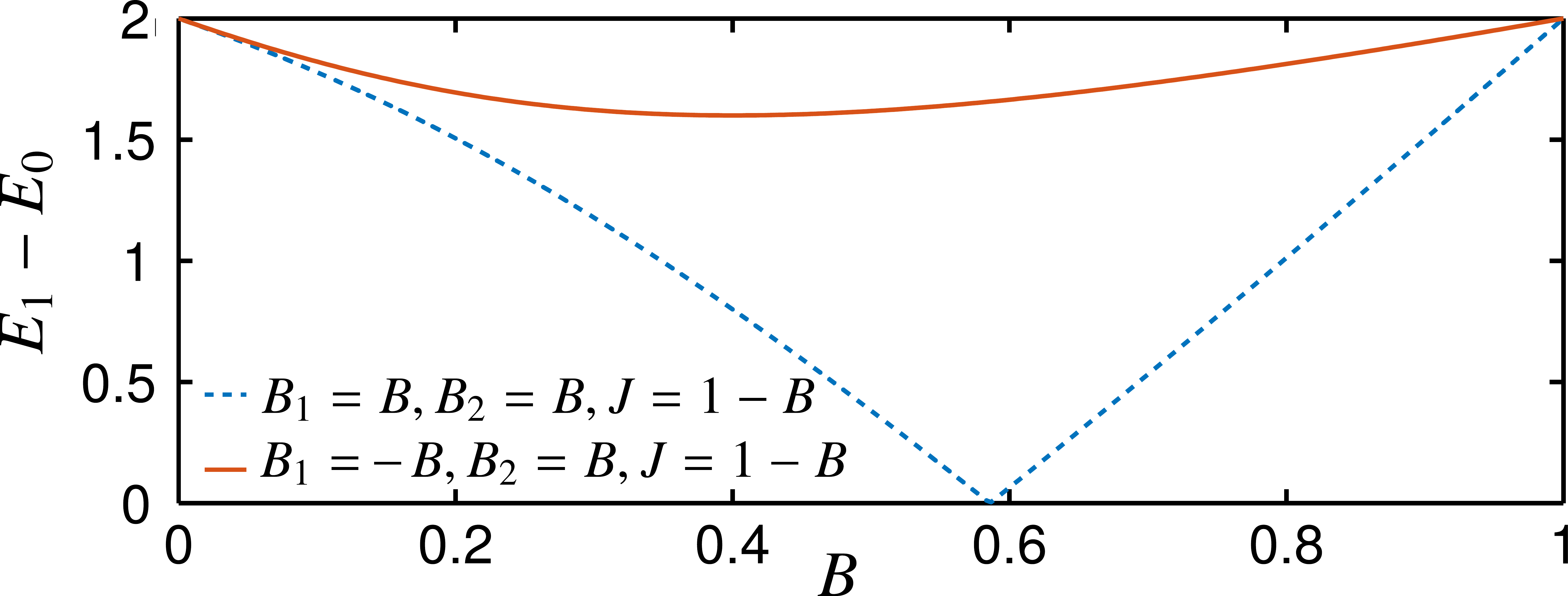}}
\caption{The energy gap as a function of $B$ (with $J=1-B$) for two cases with and without a level crossing.\label{fig:1} The units are fixed in all figures by setting both $\Lambda$ and $\hbar$ to unity.}
\end{figure}

To transform the initial states into the final target state, we need to turn off $B$ and turn on $J$. If this is done slowly enough and there is a gap to the excitations, the adiabatic theorem guarantees that the target state can be reached. Therefore, we first check the presence of a gap along a trajectory that connects the initial and final Hamiltonians. Factoring out the $B$ coefficient from the Hamiltonian, we observe that the spectrum behaves as $ E(J/B)B$. To explore all ratios $J/B$, we fix $J=1-B$ and plot the energy gap as a function of $B$ in the range $0<B<1$. As seen in Fig.~\ref{fig:1}, the gap never closes for $B_1=-B_2$ but closes at some intermediate value of $B$ for $B_1=B_2$.

In the $B_1=B_2=B$ case, the level crossing occurs at $J/B={1/ \sqrt{2}}$, implying $B=\sqrt{2}/(1+\sqrt{2})$ for $J=1-B$ at the gap closure as seen in Fig.~\ref{fig:1}.
The level crossing is exact and protected by the symmetry $|\uparrow\downarrow\rangle\leftrightarrow |\downarrow\uparrow\rangle$. In other words, the permutation operator 
\begin{equation}
Q=\left(\begin{array}{cccc}
1 & 0 & 0 & 0 \\ 
0 & 0 & 1 & 0 \\ 
0 & 1 & 0 & 0 \\ 
0& 0& 0 & 1
\end{array} \right),
\end{equation}
in the  $(|\uparrow\uparrow\rangle, |\uparrow\downarrow\rangle, |\downarrow\uparrow\rangle, |\downarrow\downarrow\rangle)$ basis, commutes with the Hamiltonian. All eigenstates of $H$ are also eigenstates of $Q$ with eigenvalues $q=\pm 1$. These two symmetry sectors are decoupled making the level crossing exact. This same symmetry forbids the transformation of $|\psi^+(0)\rangle $  to the target state  \eqref{eq:target} by any coherent nonequilibrium evolution generated by $H$.  In fact, the singlet  \eqref{eq:target} is the only eigenstate of the Hamiltonian with $q=-1$ for arbitrary $B$ and $J$ (not necessarily the ground state) and cannot be reached from any other eigenstate.
%
%

%

To see this explicitly, notice that the time dependent wave function can be written as 
\begin{equation}\label{eq:c}
|\psi(t)\rangle = c_1(t)|\uparrow\uparrow\rangle+c_2(t)|\uparrow\downarrow\rangle+c_3(t)|\downarrow\uparrow\rangle+c_4(t)|\downarrow\downarrow\rangle,
\end{equation} with the amplitudes $c_j$ evolving according to
\begin{eqnarray}
{d\over dt }c_1 (t)&=& -i\left[B_2 (t) c_2 (t)+B_1(t) c_3 (t)\right],\\
{d\over dt }c_2  (t)&=& -i[B_2 (t) c_1 (t)+2 J (t) c_3 (t)+B_1  (t)c_4 (t)],\\
{d\over dt }c_3 (t)&=& -i[B_1(t) c_1(t)+2 J(t) c_2(t)+B_2 (t)c_4(t)],\\
{d\over dt }c_4(t) &=& -i[B_1 (t)c_2(t)+B_2 (t)c_3(t)].
\end{eqnarray}
We see that in the $B_1=B_2$ case, the equations are symmetric under the $c_2\leftrightarrow c_3$ exchange. As the initial state also has $c_2=c_3$, this equality holds at arbitrary times under all possible dynamics generated by arbitrary $B(t)$ and $J(t)$ so the target state, which has $c_2=-c_3$, cannot be reached by any protocol. Hereafter, we focus on the $B_1=-B_2$ case, where the preparation of the target state is not forbidden.

\section{brute-force optimization: optimal vs linear performance}\label{sec:brute}

Our goal is to reach the target ground state~\eqref{eq:target} from the initial state~\eqref{eq:psi-} in the shortest amount of time possible by adjusting the Hamiltonian parameters as a function of time.  In real-world applications,  we may have a shorter total time than the minimum time needed to reach the state exactly. Therefore, it is useful to be able to quantify the performance of different protocols in a fixed total time $\tau$. This will also provide a practical approach for finding the minimum $\tau$ for which exact preparation is possible. Using the overlap between the final state $|\psi(\tau)\rangle$ and the target state~\eqref{eq:target}, we define the error as 
\begin{equation}\label{eq:cost1}
{\cal E}=1- |\langle \psi_{\rm target}|\psi(\tau)\rangle|^2.
\end{equation}
The error above is always nonnegative and vanishes if the two states are the same.

\begin{figure}[t]
\centerline{\includegraphics[width=.8\columnwidth]{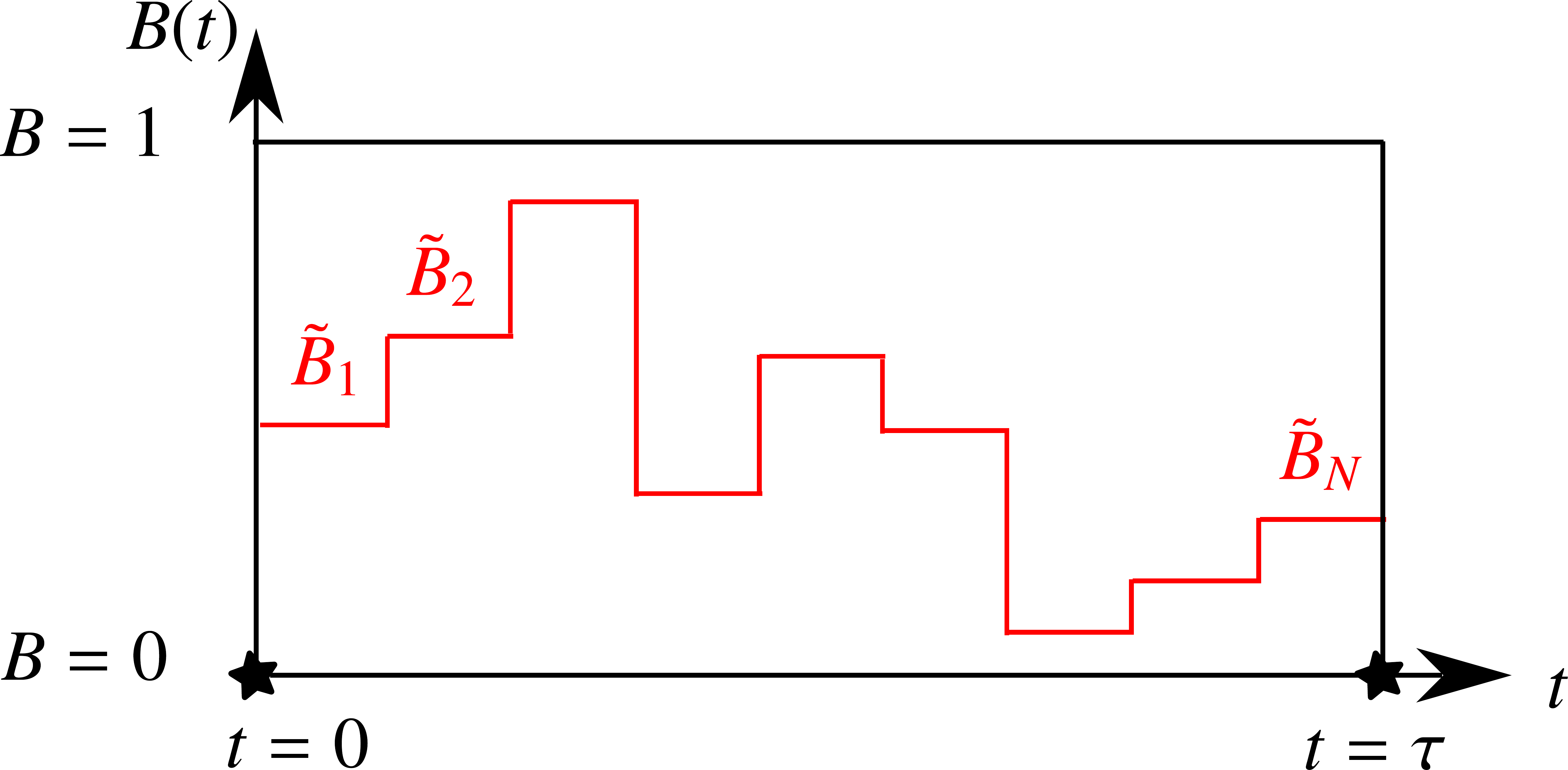}}
\caption{Approximating an arbitrary protocol with a piece-wise constant protocol to transform the functional minimization to a multivariable function minimization. The bias can be eliminated by increasing $N$ and obtaining convergent results.\label{fig:2} }
\end{figure}

The error~\eqref{eq:cost1} is a functional of the controllable time-dependent parameters $B(t)$ and $J(t)$ in the range defined in Eq.~\eqref{eq:range}. To apply standard numerical optimization algorithms, we need to transform the functional to a multivariable function. There are multiple ways to do this, e.g, using truncated coefficients of a Taylor or Fourier expansion. For our bounded parameters, it is convenient to discretize time, i.e., divide $T$ into $N$ intervals of length $T/N$, as seen in Fig.~\ref{fig:2}, creating piece-wise constant functions for $B(t)$ and $J(t)$, where $B(t)=\tilde{B}_j$ for $(j-1)\tau/N<t<j\tau/N$ and similarly for $J(t)$. Then, we can minimize ${\cal E}$ as a multivariable function of $\tilde{B}_j$ and  $\tilde{J}_j$ (with $2N$ bounded variables). To avoid an artificial bias, we increase $N$ until the results converge. In our simulations, using the interior-point minimization algorithm, we used $N=5$ and $10$ and found that the protocols and the associated errors were almost identical.

\begin{figure}[t]
\centerline{\includegraphics[width=\columnwidth]{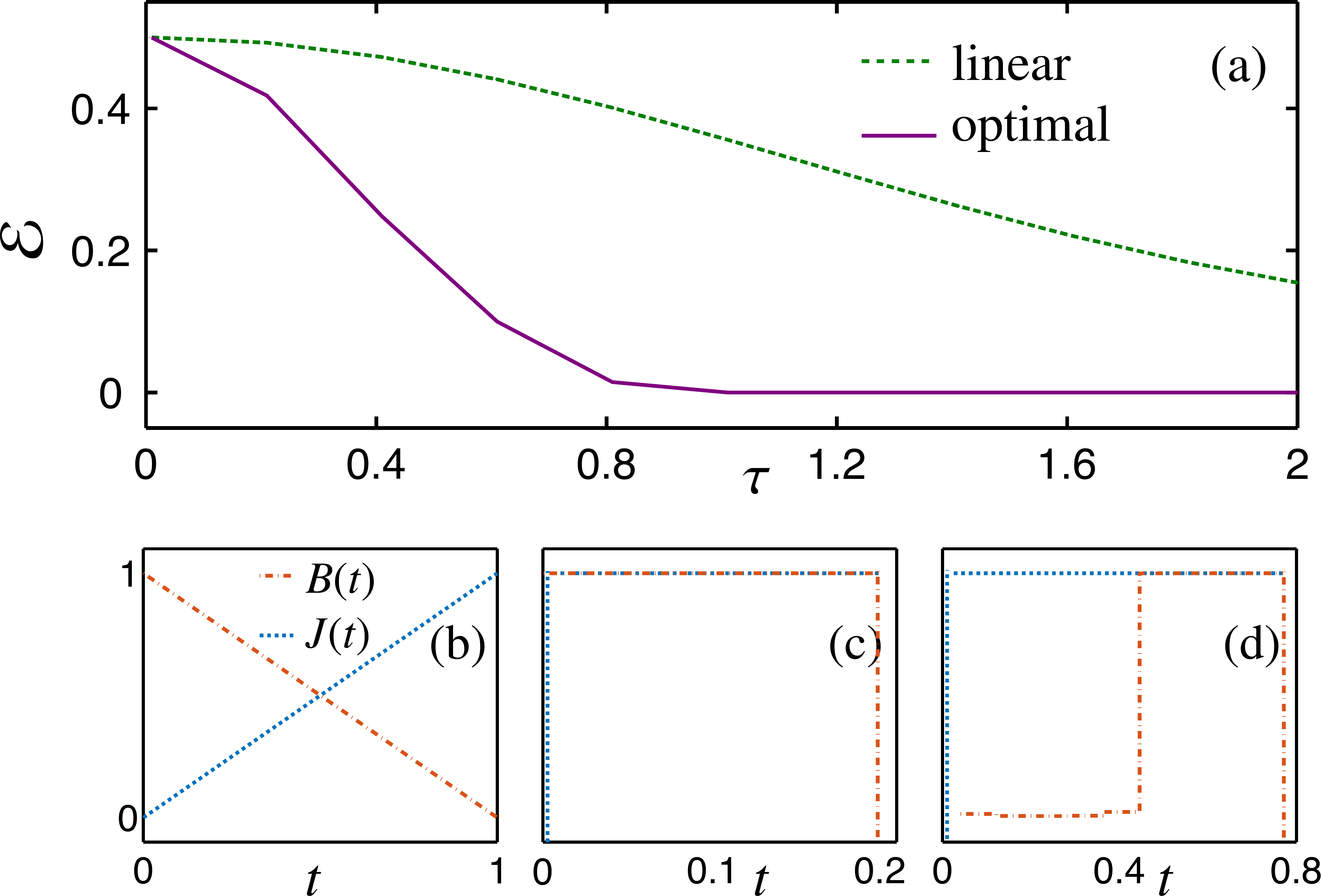}}
\caption{(a) The error for optimal and linear protocols. For a time $\tau^*\sim 0.9$, the optimal protocol is able to prepare the target entangled state exactly and the error vanishes. (b) The linear protocol for $J(t)$ and $B(t)$. (c) The optimal protocol for total time $\tau=0.2$. For $\tau<\tau_0\sim 0.4$, the optimal protocol is simply a constant pulse, with both $B$ and $J$ suddenly turned on to their maximum allowed value and kept on for $0<t<\tau$. (d) The optimal protocol for $\tau=0.8$. For $0.4\sim \tau_0<\tau<\tau^*\sim 0.9$, the optimal protocol has a constant pulse for $J$ and one switching from zero to the maximum allowed value at a finite time for $B$.\label{fig:3}}
\end{figure}

As seen in Fig. \ref{fig:3}(a), we find that our optimal protocols beat the linear protocol shown in Fig. \ref{fig:3}(b) significantly. Two examples of the optimal protocols for different values of $\tau$ are shown in  Fig. \ref{fig:3}(c) and Fig. \ref{fig:3}(d). For $\tau<\tau_0\sim 0.4$, the optimal protocol for both control parameters is simply a constant pulse, with both $B$ and $J$ at their maximum allowed value.  For $0.4\sim \tau_0<\tau<\tau^*\sim 0.9$, the optimal protocol has a constant pulse for $J$ but $B$ is initially zero for a finite time and is suddenly turned on to its maximum at a finite time $t_B$. At $\tau =0$, all protocols give an error  ${\cal E}=0.5$ (from the finite overlap of the initial and target states). Upon increasing $\tau$, the error corresponding to the linear protocol decreases, approaching ${\cal E}=0$ only at $\tau \to \infty$, while the error corresponding to the optimal one decreases more rapidly, reaching ${\cal E}=0$ at a finite time $\tau^* \sim 0.9$,
indicating an exact preparation of the desired state. This time scale is computed in units where the maximum coupling strength is set to 1 ($\Lambda=1$). The time scale is generally inversely proportional to the maximum allowed coupling strength $\tau^*\sim 0.9/\Lambda$ [see Eq. \eqref{eq:range}]. If the maximum allowed couplings are different for $B$ and $J$ but of the same order of magnitude, we still expect a time $\tau^*$, which allows for exact preparation of the target state, that is inversely proportional to the characteristic coupling strength.

We further comment on preserving the entangled state in the system after creating it with an optimal protocol that takes a time $\tau^*$. As the entangled state is the ground state for $J=1$ and $B=0$, we need to turn off $B$ at the end of the process at time $\tau^*$.
 
%
\section{Bang-bang optimization: characterizing the protocols}
\begin{figure}[t]
\centerline{\includegraphics[width=.8\columnwidth]{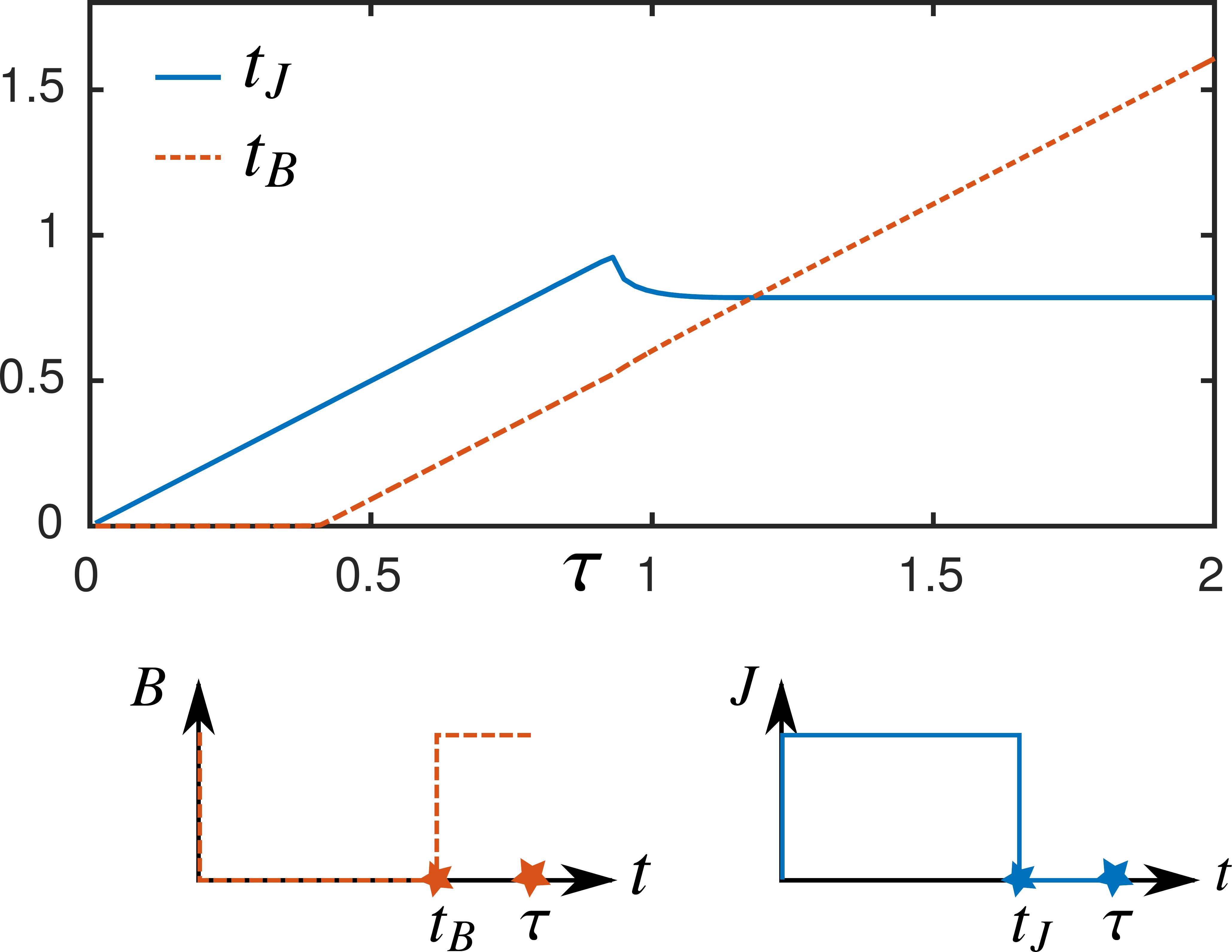}}
\caption{ The bang-bang optimal protocol and the corresponding optimal solution for $t_B$  and $t_J$ as a function of $\tau$.\label{fig:4}}
\end{figure}

From our brute-force optimization in Sec.~\ref{sec:brute}, we observe that the optimal $B(t)$ and $J(t)$ have discontinuous jumps between their minimum and maximum allowed values of zero and one. Such protocols are referred to as bang-bang protocols. As discussed in the next section, they are indeed expected to generically occur in linear optimal control problems. Knowing the bang-bang form of the protocol (and making an educated guess about the maximum number of bangs), we can perform a secondary optimization, which determines the optimal protocol very accurately. The new results are in agreement with the approximate (due to the finite discretization) results from the brute-force computations. Since we have a much smaller number of variational parameters, i.e., the times of the jumps, this optimization is much more efficient.

We performed this secondary optimization with two parameters (although one parameter would have been sufficient). As shown in Fig. \ref{fig:4}, these parameters are $t_B$ and $t_J$. $B(t)$ jumps from zero to one at $t_B$; $J(t)$ jumps from one to zero at $t_J$. We find that $t_J$ is always equal to $\tau$, so there is no jump in $J(t)$ in the middle of the evolution. As seen in the figure, for $\tau$ larger than $\tau^* \sim 0.9$, the numerically obtained $t_J$ is no longer equal to $\tau$. This is precisely the total time $\tau$ for which the optimal protocol prepares the state exactly (see Fig.~\ref{fig:3}). Therefore, for times longer than $\tau^*$, many protocols can achieve this exact preparation and the optimal protocols are not unique.  For $\tau<\tau^*$, we find two distinct behaviors for $t_B$. If $\tau$ is smaller than a critical value $\tau_0 \sim 0.4$, we have $t_B=0$. On the other hand, for $\tau_0< \tau <\tau^*$, we find the following linear relationship:
\begin{equation}\label{eq:lin_rel}
t_B=\tau-\tau_0.
\end{equation}
At this point the results above are purely numerical findings, but we will explain them in Sec. \ref{sec:pont} using the Pontryagin theorem.

\section{Connection with Pontryagin's minimum principle: singularity of the control}\label{sec:pont}
 The Pontryagin's minimum principle \cite{pontry1,pontry2} explains the bang-bang nature of the protocol above and provides an alternative approach for determining the switching time $t_B$. We first briefly review the formalism. For dynamical variables $\bf x$ evolving with the equation of motion $\dot{\bf x}={\bf f}({\bf x},{\bf b})$ with controls $\bf b$, we can write an optimal-control Hamiltonian $\cal H$ in terms of conjugate momenta $\bf p$ as
  \begin{equation}
{ \cal H}= {\bf p^{\rm T}}.{\bf f}({\bf x},{\bf b}),
 \end{equation}
where the superscript ``T'' indicates transpose. 
The dynamics of $\bf x$ and $\bf p$ are governed by the Hamiltonian equations
  \begin{equation}
\dot{\bf x}=\partial_{\bf p}{ \cal H},\quad \dot{\bf p}=-\partial_{\bf x}{ \cal H}.
 \end{equation}
 Assuming we want to minimize a cost function ${\cal E}[{\bf x}(\tau)]$ at the final time $\tau$, the boundary conditions for the conjugate momenta are given by
   \begin{equation}\label{eq:bc}
{\bf p}(\tau)=\partial_{\bf x}{\cal E}[{\bf x}(\tau)],
 \end{equation}
 and the key condition of optimal control is 
    \begin{equation}\label{eq:condition}
{\cal H}({\bf x}^{\rm opt}, {\bf p}^{\rm opt},{\bf b}^{\rm opt})=\min_{\bf b}{\cal H}({\bf x}^{\rm opt}, {\bf p}^{\rm opt},{\bf b}),
 \end{equation}
 where the superscript ``opt'' indicates the optimal protocol and the corresponding trajectories for the dynamical variables and their conjugates. A consequence of the above expression is that if the equations of motion and consequently $\cal H$ are linear in the controls, the optimal protocol is generically bang-bang.
 The only exception is the case of  \textit{singular control}, where
 the coefficient in front of a control parameter $b$ (in $\cal H$) vanishes not just at isolated points but over finite intervals. This coefficient can be written as ${ \partial {\cal H}\over \partial b}$ and thus $b$ is singular over intervals with
${ \partial {\cal H}\over \partial b}=0$.

In general, a singular optimal control parameter does not need to take its smallest or largest permissible value over such intervals.

In the context of quantum evolution, the equations of motion ${d\over dt}|\Psi\rangle=-iH(t)|\Psi\rangle$
 can be written as
 \begin{equation}
{d\over dt} {\cal R}= H(t){\cal I}, \quad {d\over dt} {\cal I}= -H(t){\cal R},
 \end{equation}
 for a \textit{real} Hamiltonian $H(t)$ (in this case a $4\times 4$ matrix), where the dynamical variables $\cal R$ and $\cal I$ contain the real and imaginary parts of the wavefunction. 
 Let us denote the conjugate momenta by vectors $P_{\cal R}$ and $P_{\cal I}$, respectively for $\cal R$ and $\cal I$. The optimal-control Hamiltonian $\cal H$ is then constructed as
 \begin{equation}\label{explicit_H}
{ \cal H}=P_{\cal R}^{\rm T} H(t) {\cal I}-P_{\cal I}^{\rm T} H(t){\cal R}.
\end{equation}
We now observe that since $H(t)$ is linear in the controls $B$ and $J$ the optimal-control Hamiltonian $\cal H$ is also linear in them. Equation \eqref{eq:condition} then implies that, at any point in time, the controls $B$ and $J$ must be set to either their minimum or their maximum allowed values depending on the sign of their corresponding coefficient in the linear function ${\cal H}$ (for optimal values of $\cal R$, $\cal I$, $P_{\cal R}$, and $P_{\cal I}$). Thus, a bang-bang solution, as found in our numerical studies, is indeed expected, unless one of the aforementioned coefficients identically vanishes over a finite time interval.

The equations of motion for the conjugate momenta are obtained by differentiating $\cal H$ with respect to the corresponding dynamical variables and are given by
 \begin{equation}
{d\over dt} P_{\cal R}= H(t)P_{\cal I}, \quad {d\over dt} P_{\cal I}= -H(t)P_{\cal R}.
 \end{equation}
 Combining the conjugate momenta into $|\Pi\rangle =P_{\cal R}+iP_{\cal I}$, we can then write
 \begin{equation}\label{eq:pi_eom}
{d\over dt}|\Pi\rangle=-iH(t)|\Pi\rangle.
\end{equation}

To proceed, we write the cost function~\eqref{eq:cost1} [see also Eq.~\eqref{eq:target}] in terms of the dynamical variables at time $\tau$ as
 \begin{equation}
{\cal E}=1-{1\over 2}\left[({\cal R}_2-{\cal R}_3)^2+
({\cal I}_2-{\cal I}_3)^2\right],
\end{equation}
where
 \begin{equation}
c_j={\cal R}_j+i{\cal I}_j, \quad j=1...4,
\end{equation}
for $c_j$ defined in Eq.\eqref{eq:c}. Using Eq.~\eqref{eq:bc} then leads to the following boundary conditions for the conjugate momenta:
 \begin{equation}\label{eq:bc2}
|\Pi(\tau)\rangle={\cal M}|\Psi(\tau)\rangle,
\end{equation}
with the matrix $\cal M$ given by
 \begin{equation}
{\cal M}=
\left(\begin{array}{cccc}
0 & 0 & 0 & 0 \\ 
0 & -1 & 1 & 0 \\ 
0 & 1 & -1 & 0 \\ 
0&0 & 0 &0
\end{array} \right).
\end{equation}

It is illuminating to use the Pontryagin equations with an ansatz characterized by one  parameter $t_B$ and $t_J=\tau$ to find the protocol shown in Fig. \ref{fig:4}. Using Eq.~\eqref{explicit_H}, we find the coefficient of $B$ in $\cal H$ (a linear function of $B$) as
 \begin{equation}
\partial_B {\cal H}=P_{\cal R}^{\rm T}{\cal K} {\cal I}-P_{\cal I}^{\rm T} {\cal K}{\cal R}={\rm Im}\langle \Pi(t)| {\cal K}|\psi(t)\rangle,
\end{equation}
where 
  \begin{equation}
{\cal K}=\partial_B H(B,J)=
\left(\begin{array}{cccc}
0 & 1 & -1 & 0 \\ 
1 & 0 & 0 & -1 \\ 
-1 & 0 &0 & 1 \\ 
0 & -1 & 1 & 0
\end{array} \right).
\end{equation}
Thus, the minimum of $\cal H$ is achieved by choosing
\begin{eqnarray}\label{eq:coeff}
B(t)&=&1,\quad -{\rm Im}\langle \Pi(t)| {\cal K}|\psi(t)\rangle >0,\\
B(t)&=&0,\quad -{\rm Im}\langle \Pi(t)| {\cal K}|\psi(t)\rangle <0,\nonumber
\end{eqnarray}

The time-dependent evolution operator can be written in a convenient form by defining 
\begin{eqnarray}
U_1(t)&=&\exp[-itH(B=0,J=1)]\equiv \exp(-itH_1),\\
U_2(t)&=&\exp[-itH(B=1,J=1)]]\equiv \exp(-itH_2),
\end{eqnarray}
which leads to
\begin{eqnarray*}
|\psi(t)\rangle&=&U_1(t)|\psi(0)\rangle,\; t<t_B,\\
|\psi(t)\rangle&=&U_2(t-t_B)U_1(t_B)|\psi(0)\rangle,\;t>t_B,\\
|\Pi(t)\rangle&=&U_2^\dagger(\tau-t){\cal M}U_2(\tau-t_B)U_1(t_B)|\psi(0)\rangle,\; t>t_B,\\
|\Pi(t)\rangle&=&U_1^\dagger(t_B-t)U_2^\dagger(\tau-t_B){\cal M}U_2(\tau-t_B)U_1(t_B)|\psi(0)\rangle,\; t<t_B
\end{eqnarray*}
where we have used the boundary condition \eqref{eq:bc2} and the equations of motion~\eqref{eq:pi_eom} for the conjugate momenta.
\begin{figure}[t]
\centerline{\includegraphics[width=\columnwidth]{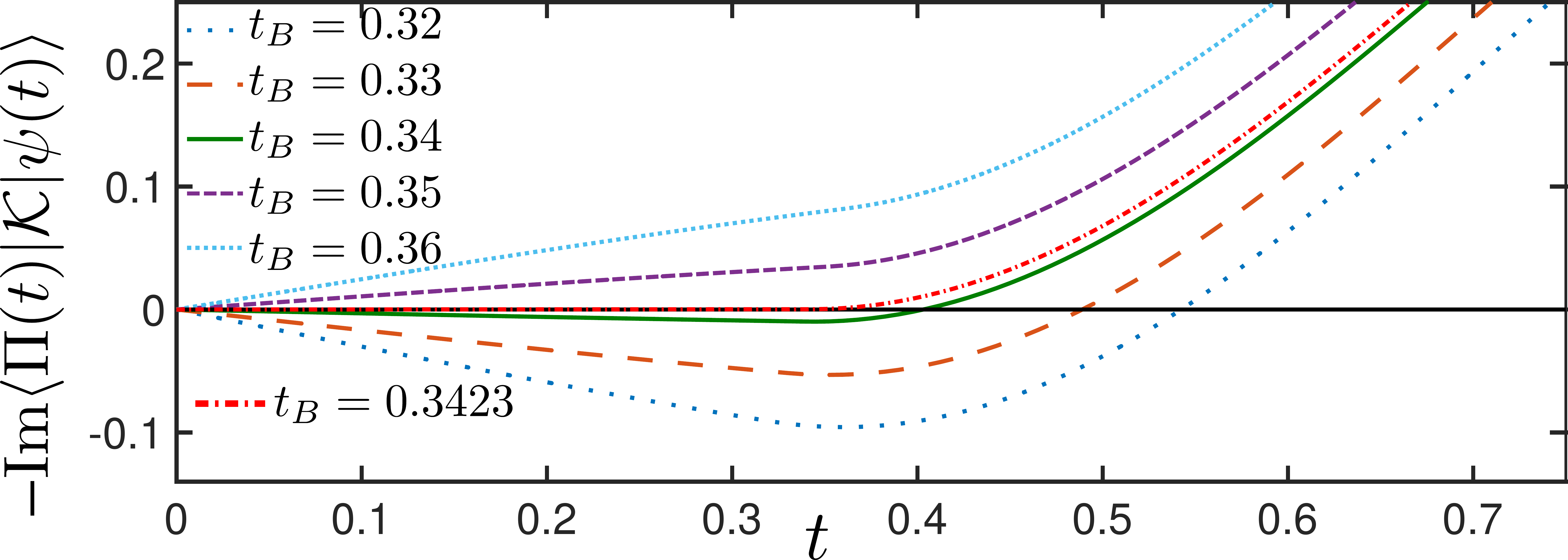}}
\caption{\label{fig:5} The factor $-{\rm Im}\langle \Pi(t)| {\cal K}|\psi(t)\rangle$ [the sign of which determines $B$ through Eq.~\eqref{eq:coeff}] as a function of time for $\tau=0.75$ and  several values of $t_B$. At the critical value of $t_B\approx 0.3423$, Eq. \eqref{eq:coeff} requires $B(t)=1$ in the entire region $t_B<t<\tau$. Interestingly, for this value of $t_B$, ${\rm Im}\langle \Pi(t)| {\cal K}|\psi(t)\rangle$ vanishes for $t<t_B$ making the control singular. Nevertheless, we have confirmed numerically that the bang-bang form with $B(t)=0$ for $t<t_B$ still provides the optimal solution.}
\end{figure}
Fixing $\tau$ and $t_B$, the four equations above uniquely determine $-{\rm Im}\langle \Pi(t)| {\cal K}|\psi(t)\rangle$ as a function of $t$. For a fixed $\tau$, we can then scan over $t_B$ and see if we can find solutions where $-{\rm Im}\langle \Pi(t)| {\cal K}|\psi(t)\rangle$ switches sign from negative to positive precisely at $t_B$. Indeed as seen in Fig.~\ref{fig:5} as an example for $\tau=0.75$ (a similar behavior was observed for other values of $\tau$), we find that for long $t_B$, $-{\rm Im}\langle \Pi(t)| {\cal K}|\psi(t)\rangle$ is positive for all finite $t$. For shorter $t_B$, there is one sign switch from negative to positive at an intermediate time. We want this switch to occur precisely at the corresponding $t_B$. Starting from the short $t_B$ limit and searching for $t_B$ for which the crossing occurs precisely at $t_B$ gives $t_B\approx 0.3423$ in agreement with our direct numerical studies, which gave $t_B=\tau-\tau_0$, with $\tau_0\sim 0.4$.

An unexpected result is that the control becomes \textit{singular} for all $t<t_B$ for our solution. This simple system thus provides an example of singular control, in which the application of the Pontryagin theorem is rather subtle. While the sign of the coefficient of $B$ determines the value of $B$, if this coefficient vanishes over a finite interval as seen in  Fig.~\ref{fig:5} for $t_B\approx 0.3423$, the control is singular and the theorem does not directly yield the optimal protocol. In case of singular control, there is no reason to expect a bang-bang protocol in intervals with a vanishing ${\rm Im}\langle \Pi(t)| {\cal K}|\psi(t)\rangle$. However, our brute-force numerical results indicate that the protocol is still bang-bang.

The numerically found relationship between $t_B$ and $\tau$ [see Fig. \ref{fig:4} and Eq. \eqref{eq:lin_rel}] in the range of $\tau$ for which the optimal control is unique and $t_B$ is finite, can be understood in terms of the Pontryagin theorem. The singularity of the control for $t<t_B$ implies ${\rm Im}({\cal C})=0$ with
\begin{equation}\label{eq:wer}
{\cal C}=\langle\psi(0)|U^\dagger_1(t_B)U^\dagger_2(\tau_0){\cal M}U_2(\tau_0)U_1(t_B)U^\dagger_1(t){\cal K}U_1(t)|\psi(0)\rangle
\end{equation}
\textit{independently} of $t$ and $t_B$ as long as we have the correct $\tau_0$. Our particular value of $\tau_0$ has the property that 
  \begin{equation}
U^\dagger_2(\tau_0){\cal M}U_2(\tau_0)=
{1\over 2}\left(\begin{array}{cccc}
-1 & e^{-i\pi/3} & -e^{-i\pi/3} & 1 \\ 
e^{i\pi/3} & -1 & 1 & -e^{i\pi/3} \\ 
-e^{i\pi/3} & 1 &-1& e^{i\pi/3} \\ 
1&-e^{-i\pi/3} & e^{-i\pi/3} & -1
\end{array} \right).
\end{equation}
Using the above matrix, Eq. \eqref{eq:wer} can be explicitly computed as a function of $t$ and $t_B$:
\begin{equation}
{\cal C}=2\cos(2t)+\cos[2(t-t_B)]-\sqrt{3}\sin[2(t-t_B)],
\end{equation}
which is a real number. Therefore, we have ${\rm Im}({\cal C})=0$ for all $t$ and $t_B$, demonstrating the validity of Eq. \eqref{eq:lin_rel}.

\section{Effects of timing error and finite bandwidth}\label{sec:error}
The bang-bang protocols require suddenly turning the control parameters on and off at precise times. For example, if we apply the protocol with $\tau=\tau^*$, we need to suddenly turn on (off) $J$ ($B$) at time $t=0$, suddenly turn on $B$ at an intermediate time $\tau^*-\tau_0$, and suddenly turn off $B$ at the end of the process at time $\tau^*$, to get a vanishing cost function. Performing a secondary optimization with the form of the optimal protocol based on the Pontryagin minimum principle gives $\tau_0=0.40774$ and $\tau^*=0.93134$.

Exact preparation of the entangled state relies on perfect timing and sharp square pulses.  Due to finite bandwidth that makes the jumps continuous or simple instrumentation inaccuracy, the applied protocol may be imprecise. We study the effects of finite bandwidth as well as timing errors by making the switching times inaccurate in a random manner. The finite bandwidth spreads out the sudden jump over a short time interval. We can expand the time-ordered exponential (appearing in the evolution operator) over this interval to first order in the duration of the interval, and generate an average coupling constant over this short interval. The same error (to leading order) can be implemented by incorrect timing for turning the coupling constant on or off.

To be explicit, we assume that instead of $t=0$ the dynamics begins at $t=\delta t_0$. Similarly, $B$ is turned on at $\tau^*-\tau_0+\delta t_1$ and turned off at time $\tau^*+\delta t_2$. We can write the cost function as
\begin{equation}
\begin{split}
{\cal E}&=1-|\langle \psi_{\rm target}|U_2(\tau_0+\delta t_2-\delta t_1)U_1(\tau^*-\tau_0+\delta  t_1-\delta t_0)|\psi(0)\rangle|^2\\
&= 1-|\langle \psi_{\rm target}|U_2(\tau_0)OU_1(\tau^*-\tau_0)|\psi(0)\rangle|^2,
\end{split}
\end{equation}
where
\begin{equation}
O\equiv U_2(\delta t_2-\delta t_1)U_1(\delta  t_1-\delta t_0).
\end{equation}

We further assume that $\delta t_j$ for $j=1,2,3$ are independent random variables, with characteristic duration $\epsilon$, and drawn from a uniform distribution $[-\epsilon/2,\epsilon/2]$. We have
\begin{equation}
\overline{\delta t_i}=0,\quad \overline{\delta t_i \delta t_j}={\epsilon^2\over 12}\delta_{ij}
\end{equation}

For small $\epsilon$, we can expand
 \begin{equation}
 \begin{split}
 O&=\openone -i\delta t_{21}H_2-i\delta t_{10}H_1 \\
 &-\delta t_{21}\delta t_{10}H_2H_1-{1\over 2}(\delta t_{21})^2H_2^2-{1\over 2}(\delta t_{10})^2 H_1^2+\dots
  \end{split}
 \end{equation}
with $\delta t_{ij}\equiv \delta t_i-\delta t_j$.
The nonvanishing averages of the above quantities are
\begin{equation}
\overline{\delta t_{21} \delta t_{21}}=\overline{\delta t_{10} \delta t_{10}}={\epsilon^2\over 6}, \quad \overline{\delta t_{21} \delta t_{10}}=-{\epsilon^2\over 12}.
\end{equation}

Up to second order in $\epsilon$, we then find the average of $\cal E$, denoted by $\overline{ {\cal E}}$, which is equal to the variation of the error from the error corresponding to the perfect protocol. (since ${\cal E}=0$ for $\delta t_j=0$). Several terms contribute to $\overline{ {\cal E}}$. We obtain
  \begin{equation}
 \begin{split}
\overline{ {\cal E}}&={\epsilon^2\over 6}{\rm Re}\left[\langle \psi(0)|{\cal U}_1^\dagger {\cal U}_2^\dagger\rho_{\rm target}{\cal U}_2(H_1^2+H_2^2){\cal U}_1|\psi(0)\rangle \right]\\
&-{\epsilon^2\over 6}{\rm Re}\left[\langle \psi(0)|{\cal U}_1^\dagger {\cal U}_2^\dagger\rho_{\rm target}{\cal U}_2H_2H_1{\cal U}_1|\psi(0)\rangle\right]\\
&+{\epsilon^2\over 6}{\rm Re}\left[\langle \psi(0)|{\cal U}_1^\dagger H_1{\cal U}_2^\dagger\rho_{\rm target}{\cal U}_2H_2{\cal U}_1|\psi(0)\rangle\right]\\
&-{\epsilon^2\over 6}\langle \psi(0)|{\cal U}_1^\dagger H_1{\cal U}_2^\dagger\rho_{\rm target}{\cal U}_2H_1{\cal U}_1|\psi(0)\rangle\\
&-{\epsilon^2\over 6}\langle \psi(0)|{\cal U}_1^\dagger H_2{\cal U}_2^\dagger\rho_{\rm target}{\cal U}_2H_2{\cal U}_1|\psi(0)\rangle +{\cal O}(\epsilon^4),\\
  \end{split}
 \end{equation}
where
\begin{equation}
\rho_{\rm target}\equiv |\psi_{\rm target}\rangle\langle\psi_{\rm target}|,\quad {\cal U}_1\equiv U_1(\tau^*-\tau_0),\quad {\cal U}_2 \equiv U_2(\tau_0).
\end{equation}
An explicit evaluation of the above expression gives
\begin{equation}\label{eq:scal}
\overline{ {\cal E}}={2\over 3}{\epsilon}^2+{\cal O}(\epsilon^4).
\end{equation}

\begin{figure}[]
\centerline{\includegraphics[width=\columnwidth]{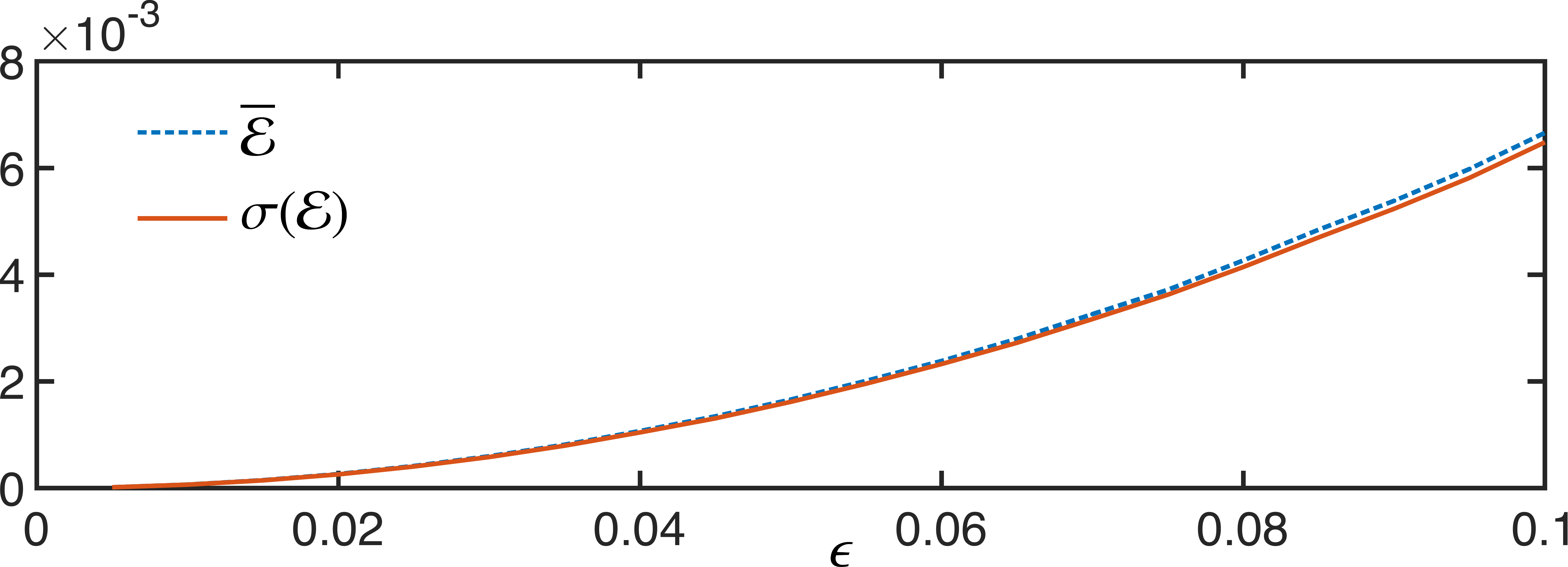}}
\caption{\label{fig:6}The average $\cal E$ and its standard deviation due to timing errors.}
\end{figure}

We verify this calculation by numerically generating many inaccurate protocols, calculating the cost function for each realization, and averaging them over the realizations. We used  $10^5$ realizations, which lead to excellent convergence. The results are shown in Fig.~\ref{fig:6} and the data are in excellent agreement with the prediction \eqref{eq:scal}. The averaged $\overline{ {\cal E}}$ does not contain enough information to determine how an individual inaccurate protocol performs. We also need to quantify the deviations of the errors for individual protocols from the above average by finding the standard deviation
\begin{equation}
\sigma({\cal E})=\sqrt{\overline{{\cal E}^2}-\left(\overline{\cal E}\right)^2}.
\end{equation}

While it is possible to determine $\sigma({\cal E})$ analytically by a similar perturbative expansion, the calculation is lengthy and not very illuminating. The same numerical simulation, however, readily yields the standard deviation of the errors. A very good fit to the data gives $\sigma({\cal E})\approx 0.647\epsilon^2$, slightly less than $\overline {\cal E}$.  Importantly, the standard deviation also scales as $\epsilon^2$. Therefore, we expect the typical errors due to imprecise implementation of the protocol to scale as quadratic in $\delta t$. Even for an $\epsilon=0.02$ (typical error of around 2\% of the total evolution time in each of the three switching times), the error is negligibly small.

\section{Conclusions}

In summary, we used optimal control to generate a maximally entangled quantum state from an unentangled state using quantum dynamics in a simple two-qubit system. The quantum dynamics was generated by a two-parameter Hamiltonian relevant to the gmon architecture of superconducting qubits. We found that when the adiabatic theorem fails due to a level crossing, the symmetry responsible for the crossing also forbids state transformations by using a more general nonadiabatic optimal protocol, making the target state unreachable. In the case of a reachable target, for various total times, we numerically found the optimal protocols that maximized the overlap of the final states with the target state. 
We found optimal protocols that substantially outperform a linear adiabatic protocol. In fact, they prepare the states exactly for a total of time of order 1. 

The optimal protocols were found to have a bang-bang character. Furthermore, we had a maximum of only one jump on one of the controls, allowing for a full characterization of the optimal solution. Taking advantage of the bang-bang form of the solution, which significantly reduces the number of the variational parameters, we then performed a much more efficient optimization. As argued in Ref.~\cite{Yang} , we expect the bang-bang ansatz to provide substantial advantages in the many-body context. We also presented an analytical understanding of the optimal protocols using the Pontryagin's minimum principle. Interestingly, we found these bang-bang optimal protocols despite the presence of singular segments in the control. Our results shed light on the conditions for reachability, and fully  characterize both qualitative and quantitative characteristics of the optimal pulses in a shortcut to the adiabatic evolution, which creates a maximally entangled state.

\acknowledgements{AR is grateful to Pedram Roushan for helpful discussions.
 This paper was supported by NSERC, Max Planck-UBC Centre for Quantum Materials, and Western Washington University. We acknowledge support from UBC Research Experience Program (REX) and Undergraduate Research Opportunities (URO).}

\end{document}